\documentclass[aps,amsmath,amssymb,prl,twocolumn,showpacs]{revtex4}

\usepackage{amsfonts}
\usepackage{amsmath}
\usepackage{amssymb}
\usepackage{graphicx}
\usepackage{color}

\newcommand{\kvec}{\mathbf{k}}
\newcommand{\avec}[1]{\mathbf{#1}}

\newcommand{\del}{\nabla}
\newcommand{\beq}{\begin{equation}}
\newcommand{\eeq}{\end{equation}}
\newcommand{\bsp}{\begin{split}}
\newcommand{\esp}{\end{split}}
\newcommand{\bpm}{\begin{pmatrix}}
\newcommand{\epm}{\end{pmatrix}}

\newcommand{\eq}[1]{Eq. [\ref{#1}]}
\newcommand{\fig}[1]{Fig.[\ref{#1}]}

\newcommand{\bib}[6]{\bibitem{#1} #2, #3, \textbf{#4}, #5, (#6).}

\begin{document}

\title{Measuring the Chern number with quantum oscillations}

\author{Anthony R. Wright}
\affiliation{School of Mathematics and Physics, University of Queensland, Brisbane, 4072 Queensland, Australia}

\date{\today}

\begin{abstract}
A peculiar feature of the majority of three dimensional topological insulator surface states studied experimentally thus far, namely their particle-hole asymmetry, makes quantum oscillations (Shubnikov de Haas and de Haas van Alphen oscillations) in these materials particularly rich. I show that this peculiarity can be exploited to measure the Chern number, and detect topological phase transitions in topological insulator surface states from the quantum spin Hall phase to the quantum anomalous Hall phase. I consider the behaviour of quantum oscillations in topological insulator thin film surface states in the presence of a topological exciton condensate, or hybridisation between the two surfaces. As a function of Zeeman field, the Chern number and phase transition from a quantum spin Hall to a quantum anomalous Hall phase can be measured using standard techniques. This effect relies necessarily on the particle-hole asymmetry which is ubiquitous in currently know materials that exhibit topological insulator surface states.
\end{abstract}


\maketitle

Three dimensional topological insulators\cite{3D, kanerev, zhangrev} have now been convincingly observed experimentally, initially through beautiful ARPES experiments showing Dirac-like band crossings at high symmetry points in the Brillouin zone \cite{hussain, hasan1, hasan}. An alternative method to detect surface states is by quantum oscillations, namely Shubnikov de Haas, and de Haas van Alphen oscillations. Shubnikov de Haas experiments \cite{qo1, qo2, natphys}, coming slightly later than ARPES, allowed complementary confirmation of the 2D nature of the surface states, as well as, it was hoped, a quantitative measurement of the Berry phase \cite{qo2, natphys, ren, veldhorst, nanoL, xiong, xiong2, natcom, ando33}. The conclusive determination of the Berry phase turned out to be unexpectedly subtle \cite{me}, and has not, to date, been accomplished.

\begin{figure}[tbp]
\centering\includegraphics[width=6cm]{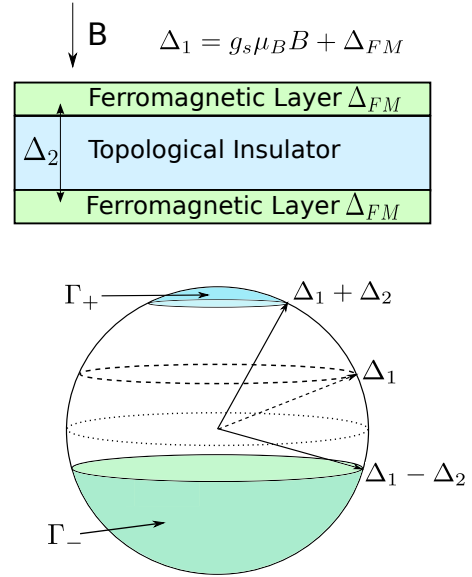}
\caption{(Color online) Upper: A topological insulator in a magnetic field, or with a ferromagnetic coating has a Zeeman bandgap, $\Delta_1 = g_s\mu_BB+\Delta_{FM}$. For an ultrathin thin film, the two surface states can hybridise by tunnel coupling, or if the two surfaces are oppositely doped, an exciton condensate can form. The band-gap from these is $\Delta_2$. From \eq{dispn}, the two bandgaps add and subtract to form two massive Dirac cones with masses $\Delta_\pm = \Delta_1\pm\Delta_2$. Lower: representing the Hamiltonian of the above system on the Bloch sphere, there are two distinct Berry phases, corresponding to $\Delta_+$ and $\Delta_-$, which are simply half the solid angle subtended by the orbit along the Fermi surface.}
\label{material}
\end{figure}

Topological insulators are characterised by their gapless surface states, which are protected from time reversal invariant perturbations \cite{kane,BHZ}. If time reversal symmetry is broken, however, a gap can be opened on the surface of a topological insulator. This can be achieved through a Zeeman field, or by coating the surface of the topological insulator with a ferromagnetic layer \cite{zhangrev}, as shown in \fig{material}. The topological insulator surface then becomes a quantum anomalous Hall insulator, so-named because it supports a single chiral edge state on each surface \cite{qah}. Experimental verification of this phase has proved elusive. 

A second gap-opening mechanism in topological insulators can occur in a thin film. The two surface states can hybridise \cite{gaptheory, gaptheory2, gaptheory3,gapexpt1,gapexpt2}, or interactions between them can lead to a non-zero excitonic order parameter \cite{moore1,moore2}, as depicted in \fig{material}. The two bandgaps -- one magnetic and one thin film induced -- can compete in an antibonding state of the topological insulator, and add in the bonding state. In this case a topological phase transition can occur, namely from the quantum anomalous Hall phase (QAH), to the quantum spin Hall phase (QSH) \cite{moore2}. This topological phase transition can be quantified by the first Chern number, which is zero in the QSH phase, and is one in the QAH phase \cite{TFT}.

In this Letter, I demonstrate (see \eq{sgn2}) that quantum oscillation experiments can measure whether a topological insulator is in the quantum spin Hall (QSH) or quantum anomalous Hall (QAH) phase, and can detect topological phase transitions between the two. Curiously, these experiments rely on the seemingly inert, yet so-far ubiquitous particle hole asymmetric spectrum of topological insulator surface states \cite{hussain, ando, shafftheory}.

\begin{figure}[tbp]
\centering\includegraphics[width=8.6cm]{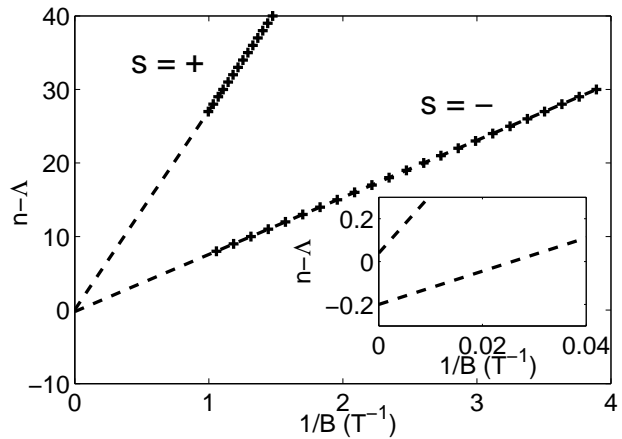}
\caption{A Landau level index plot measured over a range of Landau level filling factors (crosses), extrapolated to $1/B\rightarrow 0$ (dashed lines), measures the topological phase of a system. In this case, the two phase offsets have opposite sign (as is clear in the expanded inset), and so the system is a quantum spin Hall insulator, from \eq{sgn2}. $\Lambda = 1/2 (3/4)$ for SdH (dHvA) experiments. System parameters relevant for Bi$_2$Te$_2$Se \cite{ando}: $v_F = 3.4\times10^5ms^{-1}$, $m = 0.13m_e$, $\Delta_1 = 35$meV $ + g_s\mu_B B$, $\Delta_2 = 50$meV, $g_s = 20$.}
\label{indexfig}
\end{figure}

Shubnikov--de-Haas oscillations are the oscillations in longitudinal resistivity at external magnetic field strengths lower than the quantum Hall regime. Generically, the resistivity goes as \cite{CM}

\beq
\Delta\rho_{xx} \propto \cos\biggl[2\pi\biggl(\frac{B_0}{B} - \gamma\biggr)\biggr],
\label{resistivity}
\eeq
where $B_0$ is a measure of the area enclosed by a cyclotron orbit, and $\gamma$ is a phase offset. Semi-classically, it has been shown \cite{LK} that in two dimensions $B_0$ is well approximated by $B_0 = \frac{S(\mu)}{2\pi}$, where $S(\epsilon)$ is the area enclosed by the cyclotron orbit at constant energy $\epsilon$, and can usually be determined from the zero-field dispersion. By measuring the extrema of the resistivity or magnetisation with varying field strength, one can map the location of the filled Landau levels as a function of inverse field (though not uniquely). Extrapolating these results to $1/B \rightarrow 0$ determines the phase offset $\gamma$. Such a plot is called a Landau level index plot, a sample of which is shown in \fig{indexfig}. For normal fermions, it is well known that $\gamma = 1/2$, and for Dirac fermions (massless \emph{and} massive) one expects $\gamma = 0$ \cite{LK, sharapov}. The latter was famously observed for graphene \cite{novogeim,Zhang}, directly demonstrating its relativistic low energy spectrum.

The analysis of quantum oscillations in three dimensional topological insulators  is more nuanced than was originally expected. Specifically, the determination of the Berry phase via the intercept of the Landau level index plot as $B\rightarrow \infty$, yielded non-universal phase offsets $-1/2 < \gamma \le 1/2$ \cite{qo2, natphys, ren, veldhorst, nanoL, xiong, xiong2, natcom, ando33}, where for a Berry phase $\pi$ system, one expects $\gamma = 0$, and for Berry phase $0$, one expects to obtain $\gamma = 1/2$. This discrepancy with expected results was attributed to the Zeeman effect \cite{SWP, natphys}, the non-ideal Dirac cone \cite{ando}, and it was argued that $\gamma = 0$ will be recovered if the experiments are performed in smaller fields \cite{MS2012}, or larger fields \cite{xiong}. Recently, the expected behaviour of $\gamma$ was formulated \cite{me} within a semi-classical Lifshitz--Kosevich theory\cite{LK}, and it was shown that if, and only if, \emph{both} the material is particle-hole asymmetric, and has a bandgap, $\gamma$ becomes non-universal. This work extended on the semi-classical theory of $\gamma$ formulated in the space of particle-hole symmetric Hamiltonians, for which $\gamma$ is indeed a universal quantity \cite{Mx}. 

Although perhaps an unwanted complication for measuring the Berry phase of topological insulator surface states, the non-universality of $\gamma$ makes it an additional tool in oscillation measurements which can be utilised to experimentally determine properties of the system. In particular, in this Letter I show that $\gamma$ can be used to experimentally detect a topological phase transition between the quantum spin Hall and quantum anomalous Hall phases, or to determine the topological phase at zero or small magnetic fields. However, I stress that this can \emph{only} be accomplished if the material is particle-hole asymmetric, as it is in the majority of currently known topological insulator surface states.

\begin{figure}[tbp]
\centering\includegraphics[width=8.6cm]{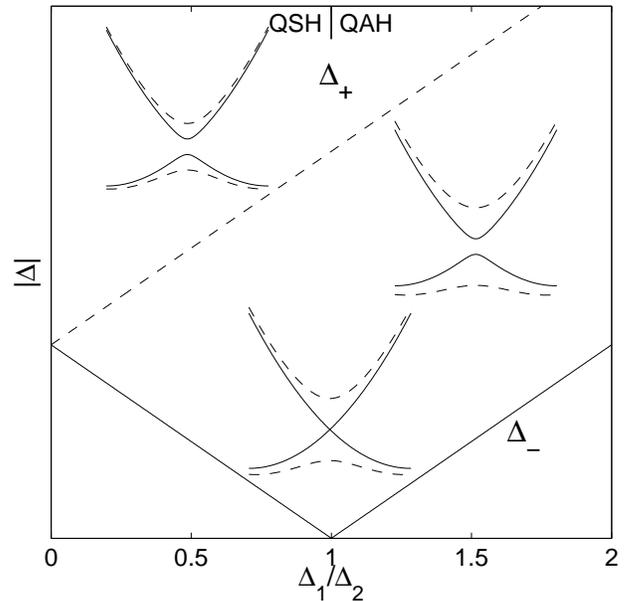}
\caption{The two bandgaps in the dispersions \eq{dispn}, add $\Delta_+$ or subtract $\Delta_-$, such that there is a critical point in one Dirac cone where its mass disappears $\Delta_-=0$, and then changes sign. The insets show the indicative dispersions of the two bands in three different regimes: $\Delta_1<\Delta_2$, $\Delta_1 = \Delta_2$, and $\Delta_1>\Delta_2$. From \eq{C}, the gap closing point marks the topological phase transition from the QSH ($\Delta_1<\Delta_2$) phase to the QAH ($\Delta_1>\Delta_2$) phase. The solid (dashed) lines correspond to $s = -(+)$, and the electron (hole) bands to $\alpha = +(-)$. }
\label{gapfig}
\end{figure}

Consider a topological insulator as shown in \fig{material}, with two surface states, intralayer ferromagnetic and/or Zeeman coupling ($\Delta_1 = g_smu_BB + \Delta_{FM}$), and interlayer exciton binding or tunnelling induced hybridisation ($\Delta_2$), with Hamiltonian matrix

\beq
H = \biggl(\frac{k^2}{2m} - \mu\biggr)\sigma_0\tau_0 + v_F\vec{k}\cdot(\vec{\sigma}\times \hat{z})\tau_z + \Delta_1\sigma_z\tau_0 + \Delta_2\sigma_0\tau_x,
\label{H}
\eeq
where $\sigma$ and $\tau$ are the Pauli matrices corresponding to spin and layer pseudo-spin, respectively. At $m\rightarrow \infty$, this is the mean-field topological exciton condensate Hamiltonian considered previously \cite{moore1,moore2}, and by a simple basis change can be written as an ultrathin film \cite{gaptheory,gaptheory2,gaptheory3}. It is clear from ARPES experiments that $m$ is finite, but can be neglected if the system is doped very close to the band-crossing point. The term $k^2/2m$ in \eq{H} explicitly breaks particle-hole symmetry, and therefore the Hamiltonian \eq{H} describes the surface of a class AII 3D topological insulator and has a $\mathcal{Z}_2$ topological invariant \cite{table}.



We can rotate the Hamiltonian \eq{H} into a block-diagonal form, consisting of two massive Dirac cones, with dispersions

\beq
\epsilon_{s,\alpha}(k) = \frac{k^2}{2m} +\alpha\sqrt{v_F^2k^2 + (\Delta_1 + s\Delta_2)^2},
\label{dispn}
\eeq
where $s,\alpha = \pm1$. So we obtain two Dirac masses $\Delta_s = \Delta_1 +s\Delta_2$ from two $2\times2$ Hamiltonians, one where $\Delta_1$ and $\Delta_2$ are competing masses, and the other where they add. The bandgaps and indicative dispersions are shown in \fig{gapfig}.

The Berry phase for a closed contour $C$ in $k-$space is

\beq
\Gamma_{s,\alpha}(C) = \oint_Cd\kvec\cdot i\langle u_{\kvec,s,\alpha}|\del_{\kvec}u_{\kvec,s,\alpha}\rangle,
\label{berry}
\eeq
where $|u\rangle$ is the eigenvector of the Hamiltonian matrix. For a $2\times2$ Hamiltonian, the Berry phase has a simple geometric interpretation on the Bloch sphere as half the solid angle enclosed by the orbit, as shown in \fig{material}. 
For our topological insulator thin film, we can readily calculate the Berry phase for contours of fixed energy $\epsilon$, giving 

\beq
\Gamma_{s,\alpha}(\epsilon) = \pi\alpha\Biggl[1 - \frac{\Delta_1 + s\Delta_2}{mv_F^2(1 + \sqrt{1 + \frac{2\epsilon}{mv_F^2} + \frac{(\Delta_1 + s\Delta_2)^2}{(mv_F^2)^2}})}\Biggr].
\eeq
The semi-classical expression for the phase offset in quantum oscillation experiments can be calculated using the bare band dispersions, together with a magnetization contribution to the band energy \cite{Niu, Mx} $\epsilon_{s,\alpha}(\kvec) = \epsilon^{B=0}_{\alpha,s}(\kvec) - \mathcal{M}_{\alpha,s}(\kvec)\cdot\mathcal{B}$.


The phase offset in quantum oscillations ($\gamma_{s,\alpha}(\epsilon)$) can now be determined, following the procedure of Ref. [\onlinecite{Mx, me}].
For our Hamiltonian \eq{H}, we obtain

\beq
\gamma_{s,\alpha}(\epsilon) =   \frac{\alpha(\Delta_1 + s\Delta_2)}{2mv_F^2\sqrt{1 + \frac{2\epsilon}{mv_F^2} + \frac{(\Delta_1 + s\Delta_2)^2}{(mv_F^2)^2}}}.
\label{gam2}
\eeq
Since the denominator in \eq{gam2} is positive definite, we can immediately state that 
\beq
\mathrm{sgn}(\gamma_{s,\alpha}) = \alpha \mathrm{sgn}(\Delta_1 + s\Delta_2).
\label{sgn}
\eeq

The first Chern number for a $2\times2$ Hamiltonian takes the particularly simple form \cite{zhangrev,moore2}

\beq
C = \frac{1}{4\pi}\int d\avec{k}\hat{d}\frac{\partial \hat{d}}{\partial k_x}\frac{\partial \hat{d}}{\partial k_y},
\label{C}
\eeq
where $H = d_0\sigma_0 + \hat{d}\cdot\vec{\sigma}$, and $\hat{d} = (d_x,d_y,d_z)/|d|$. In the case of Hamiltonian \eq{H}, $C$ becomes simply $\mathrm{sgn}(d_z)$. Combining \eq{C} with \eq{sgn} then, we obtain

\beq
\bsp
C_\alpha &= \frac{1}{2}\Bigl(\mathrm{sgn}(\gamma_{+,\alpha}) + \mathrm{sgn}(\gamma_{-,\alpha})\Bigr) = 
\left\{
\begin{array}{rl}
0, &  \mathrm{QSH} \\
1, &  \mathrm{QAH}
\end{array} \right.
\end{split}
\label{sgn2}
\eeq
\eq{sgn2} is the central result of the current work. It shows that the sign of the phase offset for a particular band is the Chern number of that band. This depends entirely on the particle hole asymmetry of the system, as can be seen by noting that in the limit of a particle-hole symmetric surface state, $m\rightarrow \infty$, the phase offset \eq{gam2} vanishes. 

\begin{figure}[tbp]
\centering\includegraphics[width=8.6cm]{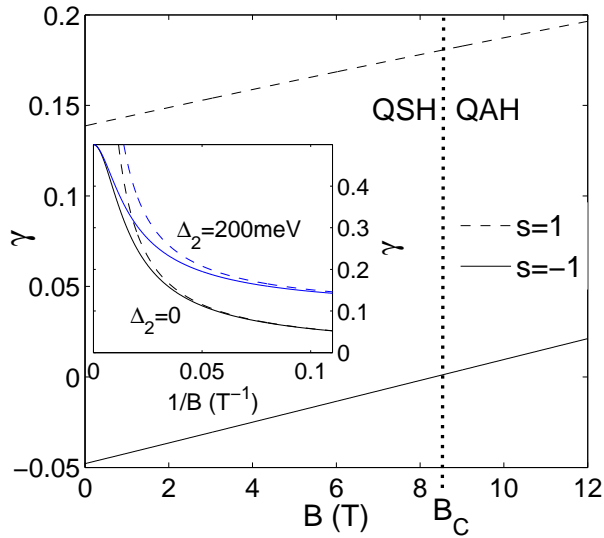}
\caption{Inducing a topological phase transition from the QSH to QAH phase by increasing the Zeeman splitting, where $\Delta_{FM} = 35$ meV and $\Delta_2 = 50$ meV. At the critical magnetic field $B_C$ (\eq{BC}), the phase offset for the $s=-1$ layer is zero. In the inset is shown the value of $\gamma(B)$ in solid lines and the estimate to $\gamma(B)$ from \eq{gammaestimate} for two interlayer gap values. For $B<10$ T, the fit is reliable. For larger fields the estimate diverges, whereas $\gamma(B\rightarrow\infty)\rightarrow1/2$. System parameters relevant for Bi$_2$Te$_2$Se \cite{ando}: $v_F = 3.4\times10^5ms^{-1}$, $m = 0.13m_e$, $g_s = 50$.}
\label{gammaB}
\end{figure}

A robust method of measuring the phase offset is to fit a nonlinear curve to the Landau level index plot \cite{me}. An example is shown in \fig{indexfig}. In the case of a gapped system at zero magnetic field, the small magnetic field expansion of \eq{gam2} can be used in the condition for extrema in \eq{resistivity} (or the corresponding result for magnetization), giving the condition

\beq
n - \Lambda \approx \frac{B_0}{B} - A_1 - A_2 B,
\eeq
where $\Lambda = 1/2\,(3/4)$ for minima in the resistivity (magnetization), $A_1 = \gamma_{B\rightarrow 0}$, and $A_2 \propto \frac{d\gamma}{dB}\Bigr|_{B\rightarrow 0}$ \cite{me}.

For a topological insulator coated with a ferromagnetic material, the analysis above can be used to determine whether in zero magnetic field, the thin film is in the QAH or QSH phase. This negates the need to measure \emph{edge} currents, or reach the quantum Hall limit. By measuring the oscillations in the longitudinal resistivity as a function of magnetic field, the Chern number can be determined.

It is also useful to know the Chern number as a function of Zeeman splitting. For instance, there is a point, when $\Delta_1 = \Delta_2$, at which a topological phase transition occurs (see \fig{gapfig}). Probing the Chern number as a function of magnetic field is, therefore, desirable. In \fig{gammaB} is shown the evolution of the phase offset $\gamma(B)$ as a function of Zeeman splitting of two topological insulator surface states with excitonic or tunnel splitting. At zero external field, the system is in the QSH phase. As the Zeeman field is increased however, there is a point when the bandgap in the $s=-1$ state closes, at which point $\gamma_{s=-1}(B) = 0$. This point marks the topological phase transition point. 

In order to measure the phase offset as a function of magnetic field, an extension to the zero field interpolation is required. At small fields we perform a Taylor's expansion on $\gamma$, which in terms of the fitting function parameters is
\beq
\gamma(B)\approx A_1 + A_2 B.
\label{gammaestimate}
\eeq
This estimate to $\gamma(B)$ allows one to determine \emph{approximately} when a topological phase transition point has been reached with increasing magnetic field. 

In the inset to \fig{gammaB}, we compare the estimate to $\gamma(B)$, \eq{gammaestimate}, with the semiclassical expression \eq{gam2} for a typical system (Bi$_2$Te$_2$Se \cite{ando}), with $g_s=50$. The fit is excellent for $B<10T$. The topological phase transition is given by $\gamma(B) = 0$. At this point, the critical field $B_C$ is reached, whereby

\beq
B_C = \frac{1}{g_s\mu_B}\bigl(\Delta_2-\Delta_{FM}).
\label{BC}
\eeq
In \fig{gammaB} is shown the critical field at which the topological phase transition occurs for a system with typical parameters.

Experiments on topological insulator thin films are rapidly improving. In particular, there now exist several experiments which have observed a hybridizing gap \cite{gapexpt1,gapexpt2}, and there are even quantum oscillation experiments\cite{ando33} on these ultrathin films (the critical width to observe hybridisation between the surfaces is six quintuple layers \cite{critical6}).

In Ref. \onlinecite{ando33}, Taskin \emph{et. al.} not only observe Shubnikov de Haas oscillations in topological insulator ultrathin films, but can also observe two separate frequencies of oscillation, characteristic of the two surfaces both contributing to the measured resistivity. With this rapid improvement in sample quality and thin film thickness control, the analyses outlined here should be attainable. Such successful experimental efforts would constitute the first direct detection of the quantum anomalous Hall phase, and would represent a valuable probe of the topological phase. 

In conclusion, I have studied quantum oscillations in topological insulator thin films with excitonic or tunnel coupling, together with Zeeman or ferromagnetic splitting. I've shown that these experiments make possible, in the case of surface states with particle hole asymmetry, the measurement of the Chern number of the sample, and thus whether it is in the quantum spin Hall phase or the quantum anomalous Hall phase. This analysis should aid the search for the elusive quantum anomalous Hall phase, and establish quantum oscillations as a robust probe of topological phases of matter.

\acknowledgments
I thank J. Kokalj and R.H. McKenzie for critical readings and insightful comments. I am financially supported by a UQ Postdoctoral Fellowship.

\end{document}